\documentclass[prb,twocolumn,floatfix,showpacs]{revtex4}
\usepackage{graphicx,psfrag}

\begin{document}

\title{ Monte Carlo Study of an Extended 3-State Potts Model on the Triangular Lattice}

\author{Z. F. Wang}
\altaffiliation[Permanent Address: ]{Physics Department, Northeast Normal University, Changchun China 130024}
\author{B.W. Southern }
\email[Email: ]{souther@cc.umanitoba.ca}
\affiliation{Department of Physics and Astronomy\\ University of Manitoba \\ Winnipeg Manitoba \\ Canada R3T 2N2}
\author{D.A. Lavis}
\affiliation{Department of Mathematics\\ King's College \\ The Strand, London \\ WC2R 2LS, UK }

\date{\today}

\begin{abstract}
By introducing a chiral term into the Hamiltonian of the 3-state Potts model on a triangular lattice additional symmetries are
achieved between the clockwise and anticlockwise antiferromagnetic states and the ferromagnetic state. This
model is investigated using  Monte Carlo methods. We investigate the full phase diagram and
find evidence for a line of tricritical points separating the ferromagnetic and antiferromagnetic phases.

\end{abstract}


\maketitle

\section{Introduction}
 The 3-state nearest neighbour ferromagnetic Potts model\cite{n7} on the triangular lattice has a second order phase transition at a finite temperature which can be determined from duality and a star-triangle relation, together with the assumption of a single critical point\cite{n11}. The critical exponents are known exactly to be {$\alpha=\frac{1}{3}, \beta=\frac{1}{9}, \gamma=\frac{13}{9}$} and { $\nu=\frac{5}{6}$}. However, the anti-ferromagnetic model displays a completely different behaviour. It is believed to have a { weak first order} transition to a six-fold degenerate ground state\cite{n5,n4,n3}. In this paper
we discuss an extension of the 3-state Potts model\cite{n6,n1} to include a chiral term which clearly exposes the first order nature of this latter transition.

\section{The Model}

The 3-state Potts model is usually defined as 
\begin{eqnarray}
{H = -J \sum_{i<j} \delta_{\sigma_i \sigma_j}}
\end{eqnarray}
where the sum is over nearest neighbours
and the variables $\sigma_i$ can have three values $A,B,C$ at each site. For the triangular lattice, it is convenient to divide the lattice into three equivalent interpenetrating sublattices corresponding to the three sites on each triangle. The Hamiltonian can then be written as a sum over {\it all} triangles $\Delta$ if nearest neighbour bonds are shared between neighbouring triangles:
\begin{eqnarray}
H &=& \sum_\Delta H_\Delta \nonumber \\
H_\Delta & =&  -\frac{J}{2} ( \delta_{\sigma_1 \sigma_2}+ \delta_{\sigma_2 \sigma_3}+ \delta_{\sigma_3 \sigma_1}) 
\end{eqnarray}
where {$\sigma_1 ,\sigma_2$ and $\sigma_3$ }correspond to the three sublattices on each triangle.

We now extend this model by adding  two additional 
three-spin terms to each triangle

\begin{eqnarray}
H_{\Delta} &=  &-\Gamma +(\Gamma -\frac{J}{2}) ( \delta_{\sigma_1 \sigma_2}+ \delta_{\sigma_2 \sigma_3}+ \delta_{\sigma_3 \sigma_1}) \nonumber \\
& & -2\Gamma \delta_{\sigma_1 \sigma_2 \sigma_3} -\Omega \epsilon_{\sigma_1 \sigma_2 \sigma_3}\end{eqnarray}
where { $ \epsilon_{\sigma_1 \sigma_2 \sigma_3}$} is the Levi-Civita symbol.  The term involving $\Omega$ is a chiral term which
distinguishes between cyclic and anti-cyclic permutations of the three states on a triangle. The  terms involving $\Gamma$ are
equivalent to adding the term $ -\Gamma \epsilon_{\sigma_1 \sigma_2 \sigma_3}^2 $ to each triangle.
Each triangle has 27 states corresponding to four different energies as indicated in Table {\ref{tab1}}.

This extended Potts model was first introduced and
studied using real space renormalization group methods by Young and Lavis\cite{n6}.  Their results agreed with previous work
by Schick and Griffiths\cite{n8} on the conventional 3-state Potts model which predicted a second order phase transition for the antiferromagnet  on this lattice. However, more recent  Monte Carlo\cite{n5,n3} results predict a weak first order transition as well as a tricritical point located in the ferromagnetic region\cite{n2,n4}. A related self-dual model which only
has the conventional three spin term on the upward triangles also exhibits a tricritical point\cite{n10}. 

In the the present work, we consider the extended Potts space which is described by the three independent coupling parameters $J, \Gamma$ and $\Omega$. The relationship of our parameters $J, \Gamma$
and $\Omega$
to the parameters $K$ and $M$ defined  by Schick and Griffiths\cite{n8} and the additional parameter $P$ defined by Young and Lavis\cite{n6}
is 
\begin{eqnarray}
K &=&(J -  2 \Gamma)/T \nonumber \\
M &=& (3J - 2 \Gamma)/T \nonumber \\
P &=& 2 \Omega /T 
\end{eqnarray}
where $T$ is the temperature.

\begin{table}
\caption{Configurations\label{tab1}}
\begin{ruledtabular}
\begin{tabular}{|c|c|c|c|}
\hline
Configuration&$\sigma_1\sigma_2\sigma_3 $&Degeneracy&$H_\Delta$\\ 
\hline
1&AAA &3& $e_1=-3J/2 \ \ \ \ \ \  $ \\ 
\hline
2&AAB & 18&$e_2=-J/2 \ \ \ \ \ \ \ $ \\  
\hline
3&ABC &3&$ e_3=-\Omega -\Gamma \ \ \ \ \ $ \\ 
\hline
4&ACB & 3&$e_4=+\Omega -\Gamma\ \ \ \ \ $  \\ 
\hline
 \end{tabular}
\end{ruledtabular}
\end{table}

\section{ Symmetries}

The model has a symmetry under the following transformations:
If we label the three states as $A,B,C$  and the three sublattices as $1,2,3$ as in Table {\ref{tab1}}, then the transformation which leaves the states on
sublattice $1$ unchanged but performs a cyclic permutation of the states on sublattice $2$ and an anti-cyclic permutation of the
states on sublattice $3$ maps the configurations on each triangle as follows:
\begin{eqnarray}
e_1 \rightarrow e_3, \ \ e_2 \rightarrow e_2,\ \  e_3 \rightarrow  e_4,\ \  e_4 \rightarrow  e_1
\end{eqnarray}
and the partition function is invariant under all permutations of the parameters $e_1, e_3$ and $ e_4$. 

A convenient set of parameters to label special points and lines in parameter space are\cite{n1}
\begin{eqnarray}
a &=& \frac{e_3- e_1}{e_2-e_1} \nonumber \\
b &=& \frac{e_4 - e_1}{e_2 - e_1}
\end{eqnarray}
The parameters $J, \Omega$ and $\Gamma$ are related to $a$ and $b$ as
\begin{eqnarray}
 \frac{\Omega}{J} &=&\frac{b-a}{2} \nonumber \\ 
\frac{\Gamma}{J} &=& \frac{3-a-b}{2}
\end{eqnarray}

{Figure 1} shows the possible ground states and the location of models corresponding to special values of $(a,b)$. Under the symmetries described above, all cases can be mapped into the upper triangular region. 
\begin{figure}
\centering
\psfrag{(Omega)}{{\large \bf \ \ \ \ \ \ ($\Omega$)}}
\psfrag{(L)}{{\large \bf ($\Gamma$)}}
\resizebox{3.5in}{!}{\includegraphics[ angle=-90]{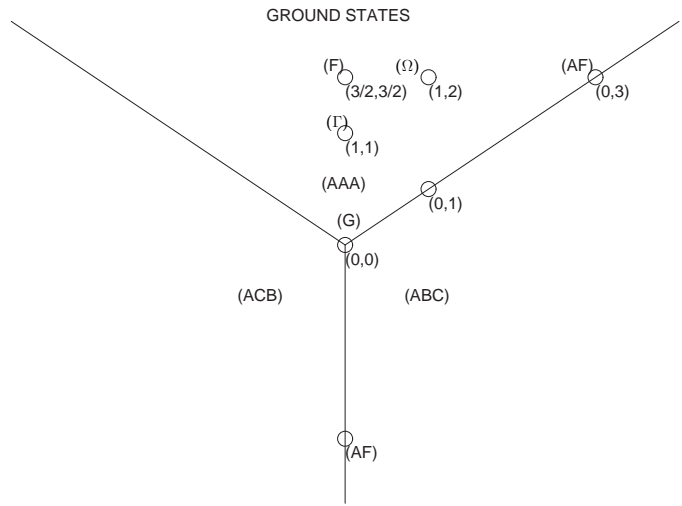}}
\caption{The possible ground states and the location of models corresponding to special values of $(a,b)$. Under the symmetries described above, all cases can be mapped into the upper triangular region.
}
\end{figure}

The point ${\sf F} (3/2,3/2)$ is the usual 3-state nearest neighbour ferromagnetic Potts model  which corresponds to  $\Omega=0, \Gamma=0$ and { ${\sf AF}$ }is the corresponding anti-ferromagnetic model which maps under the symmetries of the extended model to the point { $(0,3)$}. The point { ${\sf \Gamma} (1,1)$} corresponds  to $J=2\Gamma, \Omega=0$ and is equivalent to a 3-state Potts model with only the usual three spin term non-zero and coupling $J$. The point { ${\sf \Omega} \ (1,2)$} is an image of the model with only the chiral term { $\Omega$} non-zero. All ordered phases meet at the point {$ {\sf G} \ (0,0)$}.

\section{ Results}

The extended Potts model studied here can also be represented in terms of a spin-1 Ising model\cite{n6}. Identifying the 3 states $A,B,C$ as the spin states $S_i = \pm 1, 0$ at each site $i$, we choose the following order parameter 
\begin{eqnarray}
m= |m_1| + |m_2| + |m_3|
\end{eqnarray}
where $m_\alpha, \alpha=1,2,3$ are the sublattice spin magnetizations. At high temperatures, all three states have equal weights
and the sublattice magnetizations are zero apart from finite size effects. At low temperatures, the sublattices become ordered into
one of the three possible states. In the ferromagnetic phase, the sublattices have the same state. However in the anitferromagnetic regions, there can be either a cyclic or anti-cyclic arrangement.  We allow for the permutation symmetry of each of these phases by summing $m$ over the permutations.  In this way $m$ has the value unity in each of the ordered phases at zero temperature.
We also calculate the susceptibility $\chi$ and the Binder cumulant $Um$ associated with fluctuations in this order parameter.

We have used a standard Metropolis Monte Carlo algorithm to study lattices with $N=L^2$ sites for values of $L$ ranging from  18 to 96
 along the line $\Gamma=0$
joining (3/2,3/2) to (0,3),  followed by the line $\Omega + \Gamma = 3J/2$ from (0,3) to (0,0) and finally back along the line $\Omega=0$ from
(0,0) to (3/2,3/2).  We present our results for several points along these lines:

\begin{center}
Point ${\sf \Omega}$(1,2)
\end{center}
This point is equivalent under the permutation symmetry to the case where only the parameter { $\Omega$} is non-zero. The system  is expected to exhibit a second order phase transition\cite{n9} belonging to the same universality class as the ferromagnetic 3-state Potts model. Our Monte Carlo results are consistent with this behaviour. We have calculated the order parameter ($m$), the susceptibility($\chi$) and the Binder cumulant($Um$) for lattice sizes ranging from $L=12$ to $L=48$ where $N=L^2$ is the number of sites. Figure 2 shows the Binder cumulant $Um$ from which the crossing point of the curves for different $L$ yields an estimate of the critical temperature. Finite size scaling is then used to collapse the data for both the order parameter  and the susceptibility (figure 3) in order to determine the critical exponents $\beta, \gamma$ and $\nu$. We find $\beta=0.11 \pm 0.01, \gamma=1.44 \pm 0.01$ and $\nu = 0.82 \pm 0.06$ which are consistent with this model belonging to the same universality class
as the ferromagnetic 3-state Potts model.

\begin{figure}
\centering
\includegraphics[height=2.5in, angle=0]{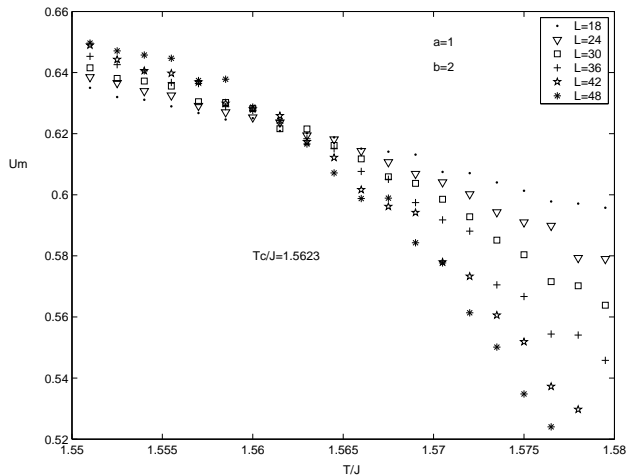}
\caption{Binder cumulant as a function of $T/J$ for sizes $L=18,24,30,36,42,48$ in the case of $a=1, b=2$. The curves cross at $T_c/J = 
1.5623$ indicating a continuous transition.}
\end{figure}

\begin{figure}
\centering
\includegraphics[height=2.5in, angle=0]{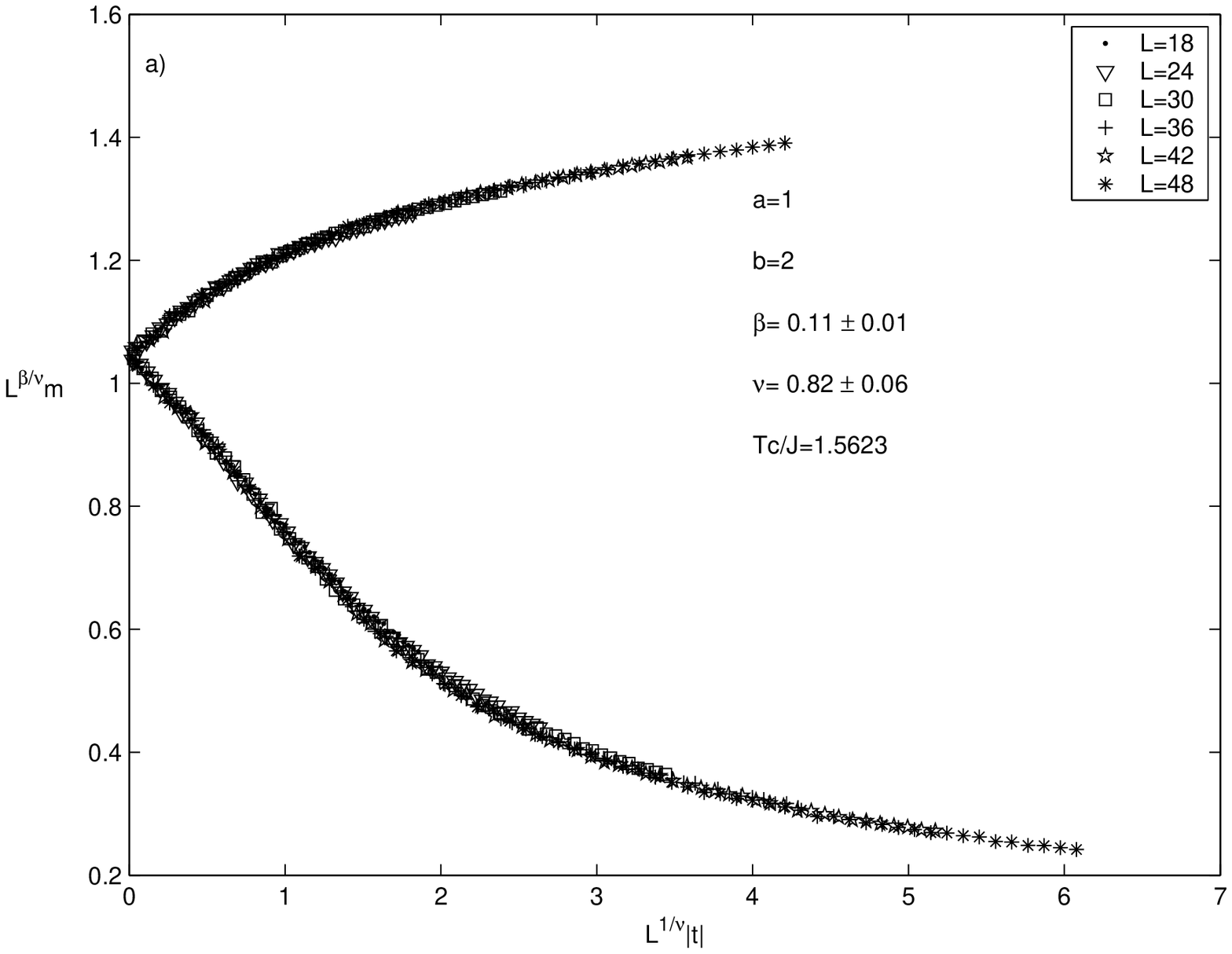}
\includegraphics[height=2.5in, angle=0]{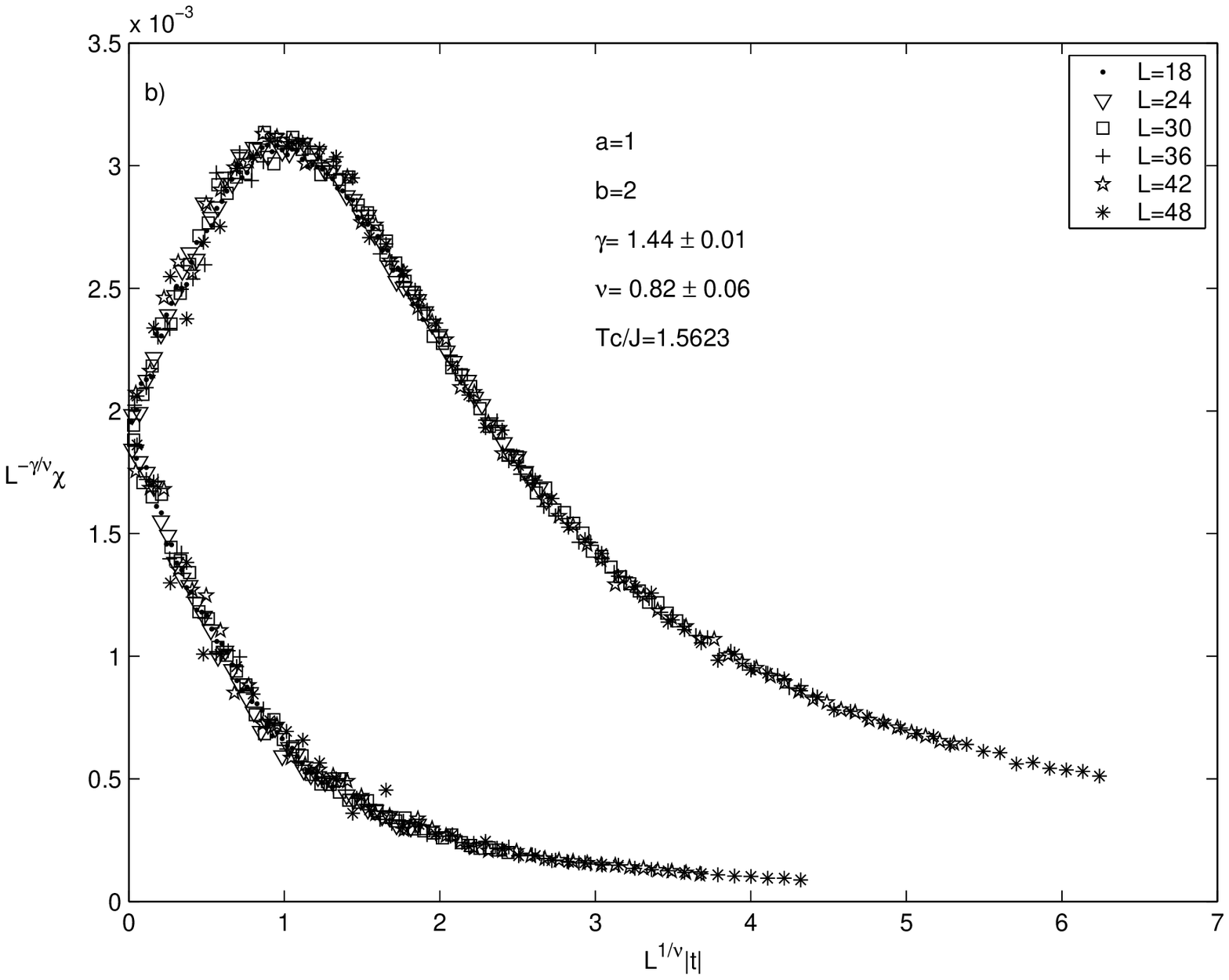}
\caption{Finite size scaling plots for the order parameter and susceptibility for different lattice sizes $L$ and reduced temperatures $t=\frac{T-T_c}{T_c}$ in the case $a=1,b=2$. a) The order parameter $m$ near $T_c$ yields the values $\beta=0.11 \pm 0.01$ and $\nu=0.82 \pm 0.06$.  b) The susceptibility $\chi$ yields the values $\gamma=1.44 \pm 0.01$ and $\nu=0.82 \pm 0.06$.}
\end{figure}

\begin{center}
{Point {\sf AF} (0,3)}
\end{center}
This point is equivalent under the permutation symmetry to the antiferromagnetic 3-state Potts model and is believed to exhibit a first order thermal phase transition. Our Monte Carlo results agree with those of Alder et. al.\cite{n3}. Figure 4a shows a histogram for the energy per site, $e =E/N$, at $T_c$ in the case of an $84 \times 84$ lattice and figure 4b shows how the energies of the two peaks depend on lattice size. The results indicate a finite jump in the energy
in the limit of an infinite lattice. The temperature and energy scales at the point $(0,3)$ are twice as large as at the equivalent {\sf AF}
point in the lower part of figure 1.

\begin{figure}
\centering
\includegraphics[height=2.5in, angle=0]{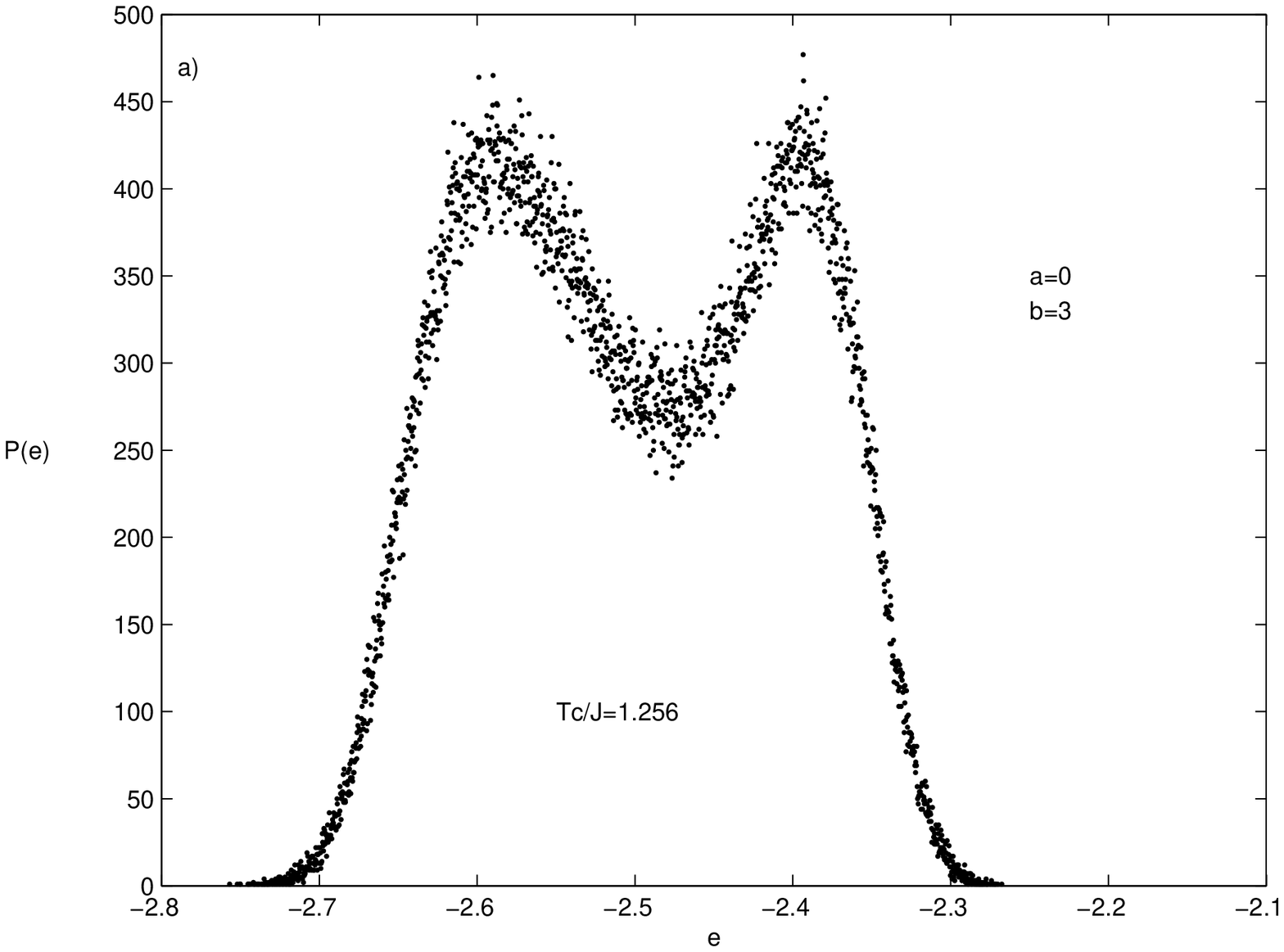}
\includegraphics[height=2.5in, angle=0]{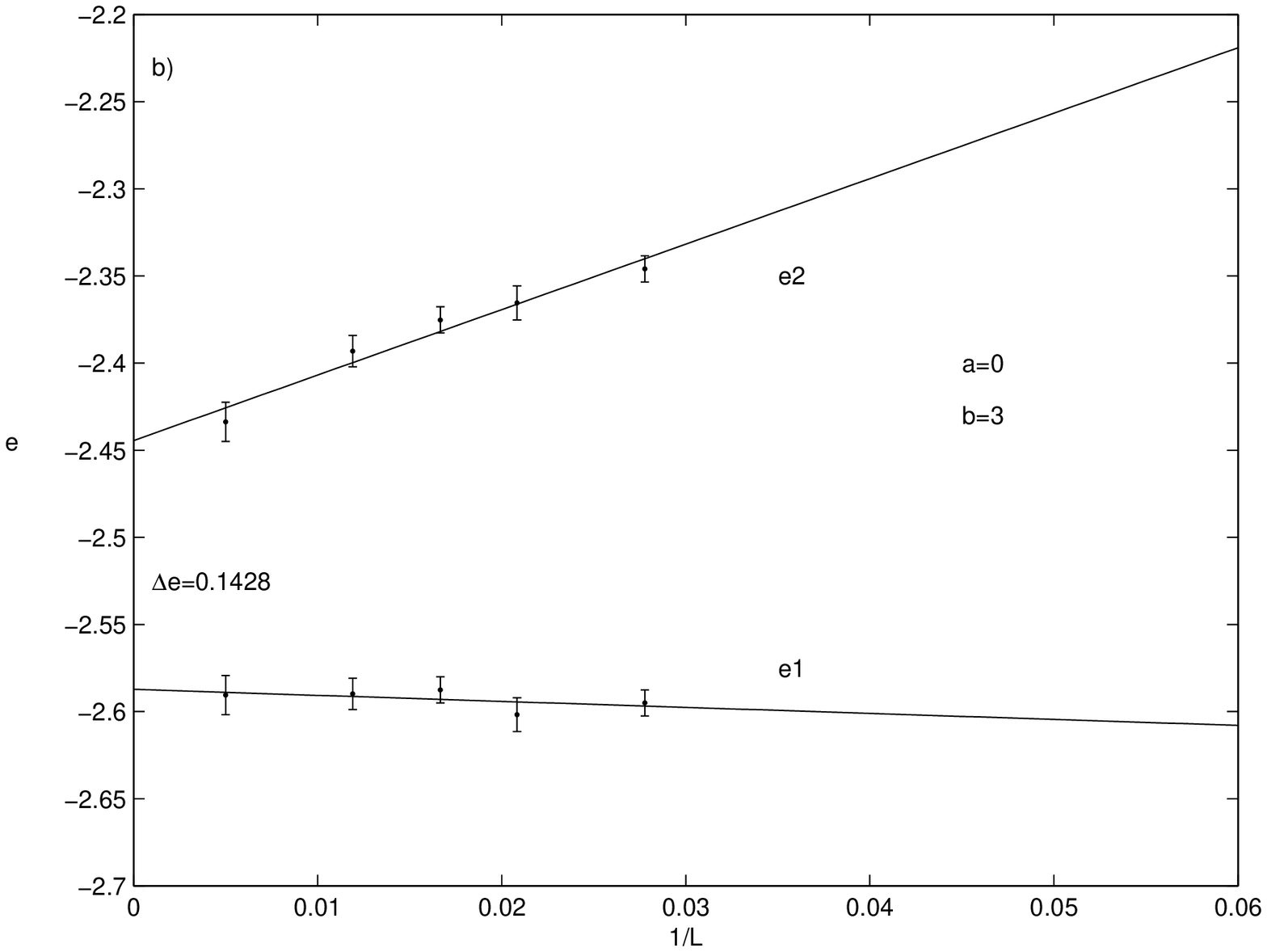}
\caption{a) Histogram for the energy per site, $e=E/N$, at $T_c/J = 1.256$ in the case $a=0,b=3$ which is equivalent to the 3-state Potts
antiferromagnet. b) Dependence of the histogram peak energies on lattice size. Extrapolation to the large $L$ limit indicates a finite jump
in the energy at $T_c$.}
\end{figure}

We have also studied this latter point directly with $\Omega=\Gamma=0$ and $J < 0$. A histogram of the chirality order parameter
$\epsilon_{\sigma_1 \sigma_2 \sigma_3}$ for an $18 \times 18$ lattice clearly indicates the first order nature of the transition.
Figure 5 shows the histogram at a few temperatures near $T_c$ which, as mentioned above,  has one half the value of the equivalent point at (0,3). At temperatures
above $T_c$, there is a single peak in the distribution at $\varepsilon=0$ but, as the temperature is lowered, two additional side
peaks appear at finite non-zero values of $\varepsilon$. At $T_c$ all three peaks have equal weight and below $T_c$ the two side
peaks dominate. This behaviour clearly indicates that the phase transition is first order. A continuous transition would correspond to the central peak splitting into two peaks which move continuously away from zero.

\begin{figure}
\centering
\includegraphics[height=2.5in, angle=0]{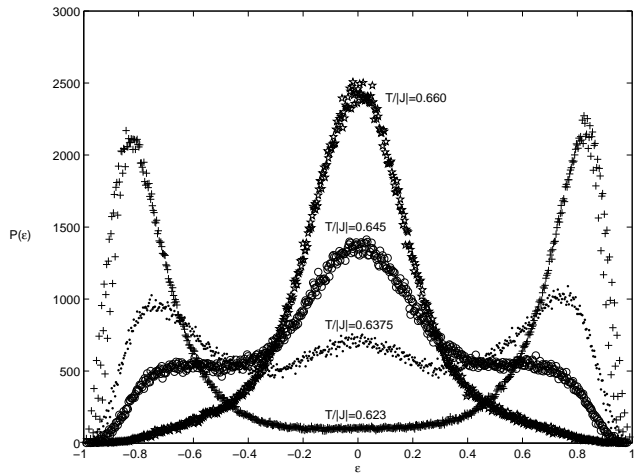}
\caption{Histogram for the chirality $\varepsilon$ at various temperatures near  $T_c$ at the {\sf AF} point with $J < 0$.}
\end{figure}

\begin{center}
{ Point (0,1)}
\end{center}
This point  lies on the first order surface separating the $AAA$ and $ABC$ ground states.  We have calculated histograms for both the energy
and the order parameter $m$ for various lattice sizes $L$ at the corresponding $T_c$. Both distributions show a double peaked structure as $T_c$ is approached. Figure 6 shows how the peak energies  and magnetizations vary with $L$. Again we find a first order thermal phase transition which is stronger than at the {\sf AF} point.

\begin{figure}
\centering
\includegraphics[height=2.65in, angle=0]{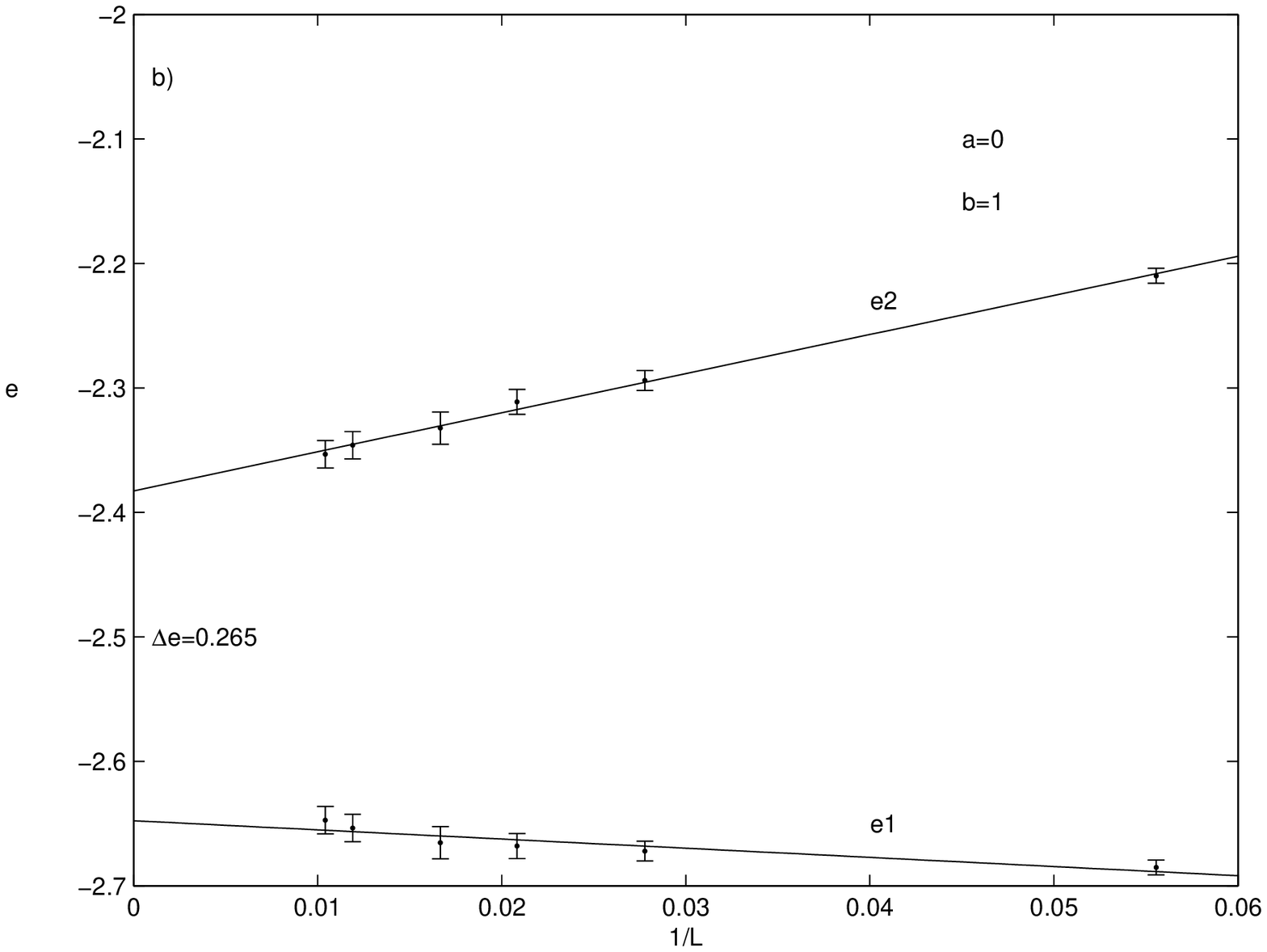}
\includegraphics[height=2.5in, angle=0]{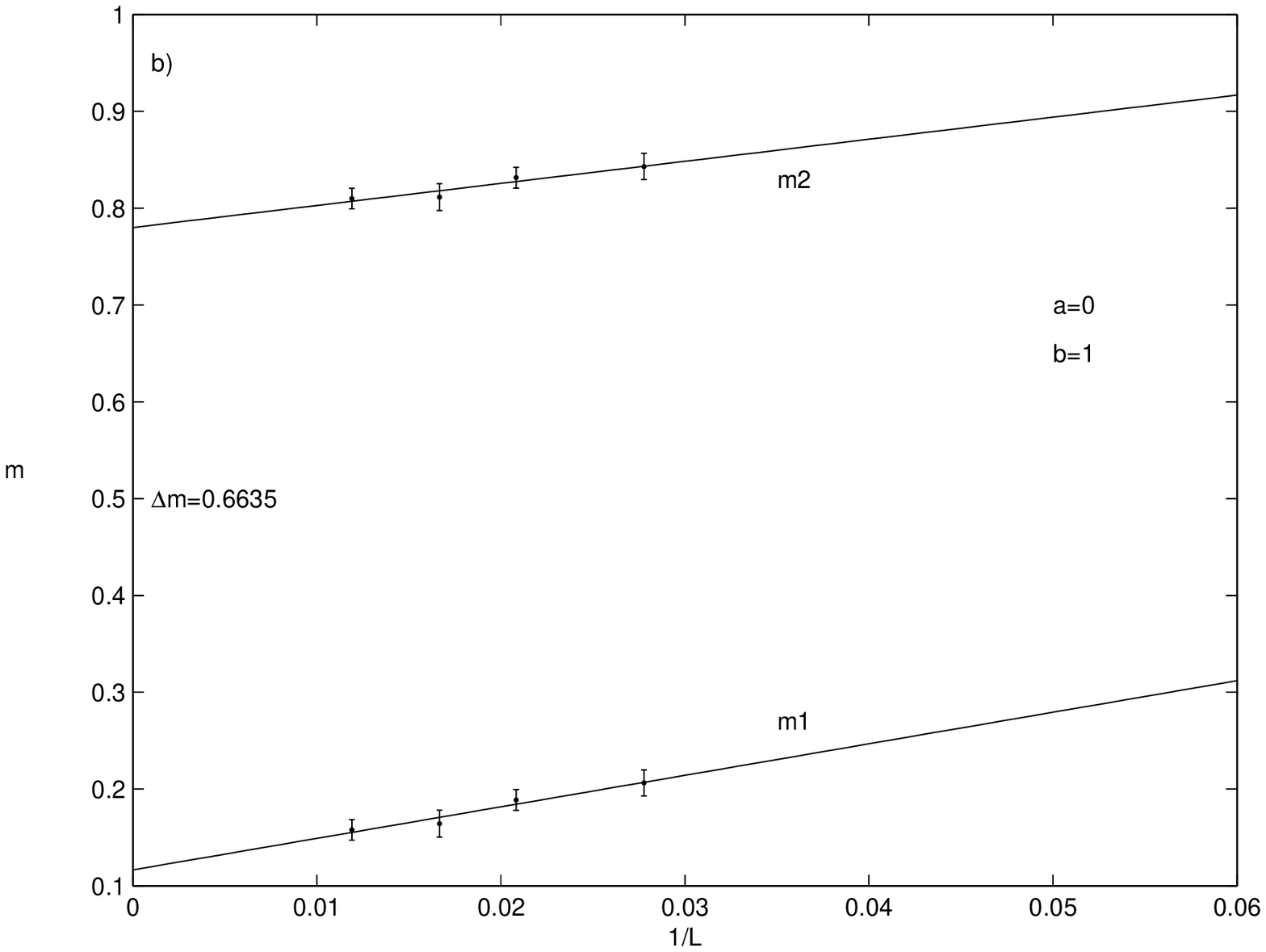}
\caption{ Dependence of the histogram peak locations on lattice size for the case $a=0,b=1$. a)   Dependence of the histogram peak energies on lattice size.   b) Dependence of the histogram peak magnetizations on lattice size. Extrapolation to the large $L$ limit indicates a finite jump
in the both the energy and the order parameter at $T_c$. }
\end{figure}

\newpage
\begin{center}
{ Point  {\sf G}(0,0)}
\end{center}
This point corresponds to a meeting of the three first order planes separating the $AAA$, $ABC$ and $ACB$ ground states. Figure 7 show a magnetization histogram for a $48 \times 48$ lattice at $T_c$ .   The jump in the magnetization is larger than at the point (0,1).
We have also studied the behaviour at a large value of $b=10 $ along the line $a=0$  and again find a first order transition which is
weaker than at the {\sf AF} point. 

\begin{figure}
\centering
\includegraphics[height=2.5in, angle=0]{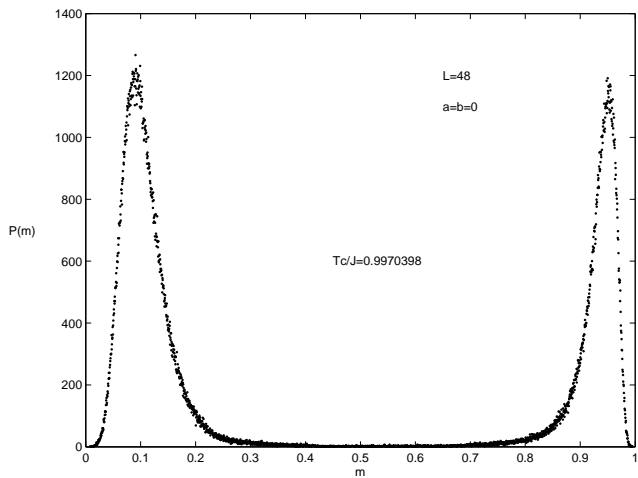}
\caption{Magnetization histogram at the point $G$ (0,0) for a $48 \times 48$ lattice at $T_c$. The peaks of the distribution
are more widely separated than at the point (0,1).}
\end{figure}

We have also investigated the possibility of a tricritical point along the line $a=b$ using the histograms. Figure 8 shows the
energy discontinuity as a function of $a$. The results were obtained by calculating histograms for various lattice sizes and extrapolating to the large $L$ limit.  The energy jump decreases as we move upward from the point ${\sf G} \  (0,0)$ . A tricritical point exists very close to $a=b=0.35$ which is in between the values
reported by the series\cite{n2} and previous Monte Carlo results\cite{n4}.
\begin{figure}
\centering
\includegraphics[height=2.5in, angle=0]{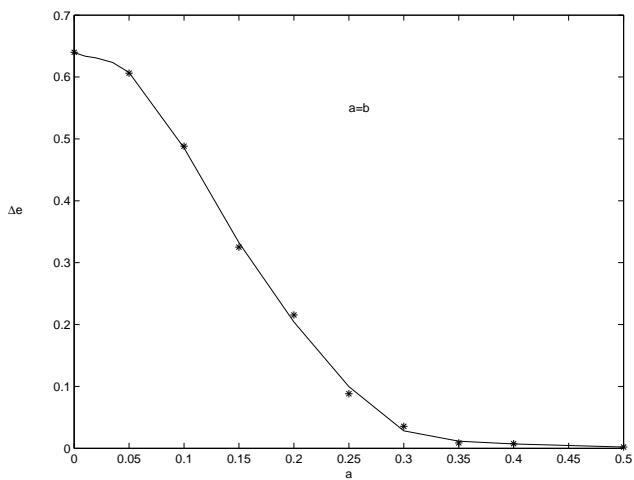}
\caption{The energy jump as a function of $a$ along the line $a=b$. A tricritical point exists near $a=b=0.35$.}
\end{figure}

{Figure 9} shows the phase diagram  along the closed path $(3/2,3/2) \rightarrow (0,3) \rightarrow (0,0) \rightarrow (3/2,3/2)$ obtained using our Monte Carlo method. Along the first section of the path, the second order transition changes to a first order transition at the point labelled ${\sf T}$ which is quite close to the ${\sf AF}$ point. As we then move towards the point ${\sf G} \  (0,0)$ along the line $a=0$, the transition remains first order with the transition temperature decreasing and the size of the energy and order parameter discontinuities increasing. Finally, along the section of the path moving back to the point ${\sf F }$ the first order transition again changes to a second order transition near the point $a=b=.35$. The points labelled by ${\sf T}$ are tricritical points.
\begin{figure}
\centering
\includegraphics[height=2.5in, angle=0]{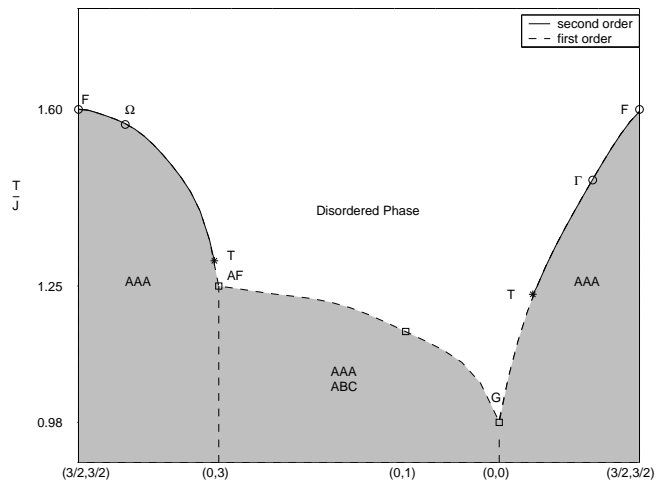}
\caption{The phase diagram  along the closed path $\Gamma=0 \rightarrow \Omega+\Gamma=3J/2 \rightarrow \Omega=0$ obtained using a heat-bath Monte Carlo method. Solid lines indicate second order transitions and dashed lines indicate first order transitions. Tricritical
points are indicated by the letter ${\sf T}$.}
\end{figure}

Figure 10 shows a schematic phase diagram in the $a-b$ plane where shaded areas correspond to first order transitions.
The shaded region narrows appreciably near the $AF$ point which is consistent with a weak first order transition.
\begin{figure}
\centering
\includegraphics[height=2.5in, angle=0]{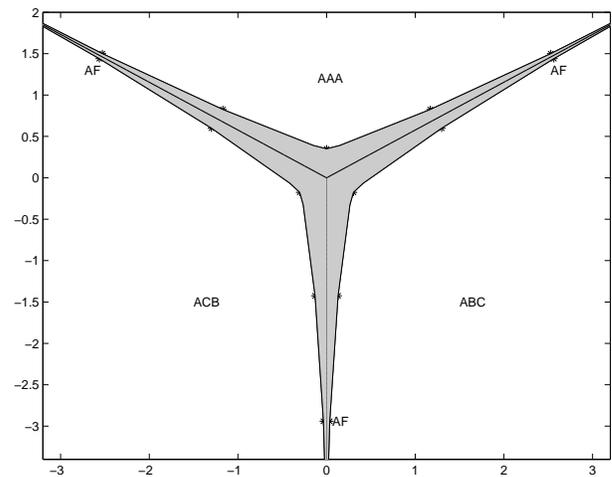}
\caption{A schematic picture of the phase diagram in the extended Potts space. The shaded area is a region of first order
thermal phase transitions. The region narrows substantially near the AF point indicating a weak first order transition.}
\end{figure}

\section{Summary}

 We have studied an extended 3-state Potts model on the triangular lattice using Monte Carlo methods. The extended model
includes a chiral term which distinguishes between cyclic and anti-cyclic configurations on each triangle and clearly exposes the first order nature of the nearest-neighbour anti-ferromagnet.
 
In the regions corresponding to continuous transitions, the model belongs to the 3-state ferromagnetic Potts model universality class.
We have not been able to locate the tricritical line with sufficient precision to determine the tricritical exponents.

\begin{acknowledgments}
This work was supported by the Natural Sciences and Research Council of Canada, the China Scholarship Council
and the High Performace Computing facility at the University of Manitoba.  DAL wishes to thank the Department
of Physics and Astronomy for its hospitality during a short visit funded by the Royal Society of London, the University
of Manitoba Visiting Professional Associate Support Program and the Winnipeg Institute for Theoretical Physics.

\end{acknowledgments}

\bibliography{zhoufei}

\end{document}